\documentclass{ws-procs975x65}

\usepackage{amsmath, amssymb}

\usepackage[draft]{epsfig}

\begin{document}

\title{Testing the paradigm of adiabaticity
\footnote{\uppercase{W}ork partially supported by the
\uppercase{S}wiss \uppercase{N}ational \uppercase{S}cience
\uppercase{F}oundation, the \uppercase{S}chmidheiny
\uppercase{F}oundation and the \uppercase{E}uropean
\uppercase{N}etwork \uppercase{CMBNET}.}}

\author{Roberto Trotta ~\ \lowercase{and}~ Ruth Durrer }

\address{D\'epartement de Physique Th\'eorique,  Universit\'e
de Gen\`eve \\ 24 quai Ernest Ansermet, CH--1211 Gen\`eve 4
(Switzerland) \\ E-mail: \texttt{roberto.trotta@physics.unige.ch}}

\maketitle

\abstracts{We introduce the concepts of adiabatic (curvature) and
isocurvature (entropy) cosmological perturbations and present
their relevance for parameter estimation from cosmic microwave
background anisotropies data. We emphasize that, while present-day
data are in excellent agreement with pure adiabaticity,
subdominant isocurvature contributions cannot be ruled out. We
discuss model independent constraints on the isocurvature
contribution. Finally, we argue that the Planck satellite will be able
to do precision cosmology even if the assumption of adiabaticity is relaxed.}

\newcommand{\ETAL}{{\it et al.}}
\newcommand{\EG}{e.g.~}
\newcommand{\IE}{i.e.~}
\newcommand{\CF}{cf}

\newcommand{\CRIT}{{\rm crit}}
\newcommand{\TOT}{{\rm tot}}
\newcommand{\MAT}{{\rm m}}
\newcommand{\CDM}{{\rm c}}
\newcommand{\BAR}{{\rm b}}
\newcommand{\FG}[1]{Fig.~#1}

\newcommand{\Om}{\Omega}
\newcommand{\La}{\Lambda}
\newcommand{\pert}[1]{\frac{\delta \rho_{#1}}{\rho_{#1}}}

\newcommand{\UUNIT}[2]{
{\;\mathrm{#1}^{#2}} }

\newcommand{\MIX}{{\rm MIX}}
\newcommand{\AD}{{\rm AD}}
\newcommand{\BI}{{\rm BI}}
\newcommand{\CI}{{\rm CI}}
\newcommand{\NID}{{\rm NID}}
\newcommand{\NIV}{{\rm NIV}}

\newcommand{\be}{\begin{equation}}
\newcommand{\ee}{\end{equation}}
\def\gsim{\raise 0.4ex\hbox{$>$}\kern -0.7em\lower 0.62
ex\hbox{$\sim$}}
\def\lsim{\raise 0.4ex\hbox{$<$}\kern -0.8em\lower 0.62
ex\hbox{$\sim$}}

\section{Introduction}

The cosmic microwave background (CMB) constitutes one of the
pillars of modern cosmology. Since the first measurement of
primary anisotropies by the COBE satellite\cite{Bennett:96} in
1992, the steady increasing precision has culminated with the WMAP
results\cite{WMAP}.

The dependence of the CMB power spectrum on cosmological
parameters is well understood, and fast numerical codes produce a
theoretical power spectrum in a few seconds with better then
percent accuracy. Therefore the extraction of cosmological
parameter by grid or Monte Carlo techniques is now a well
established practice. By ``cosmological parameters'' we mean a set
of 5 dimensionless numbers which describe the matter content of
the universe today ($\Om_\La, \Om_c,\Om_b, \Om_r, \Om_\kappa$
parameterizing the energy density in terms of the cosmological
constant, cold dark matter (CDM), baryons, radiation and
curvature, respectively), supplemented with the value of the
Hubble parameter today, $H_0$, and the optical depth to
reionization, $\tau$. In the  simplest scenario which contains
scalar perturbations only, we need two other parameters to
describe the amplitude ($A_s$) and spectral dependence ($n_s$) of
the initial perturbation spectrum. Certain combination of those
parameters constitute ``orthogonal sets'' with respect to the CMB
data, hence can be determined with very high
accuracy\cite{Kosowsky:02}.

Combination of CMB data with other cosmological data sets allows
to constrain the above 9 standard parameters within a few percent.
This is a spectacular achievement, even more so since many totally
independent measurements seem to be converging toward the same
values\cite{Tegmark:03}. The accuracy of parameter extraction
relies on the assumption that the initial conditions (IC) for the
perturbations are of the simplest possible type, namely purely
adiabatic. It is much more difficult to use the CMB to constrain
cosmological parameters and at the same time to learn more about
possible deviations from adiabaticity in the IC. Nevertheless the
CMB represents the most promising data set to learn  about the
type of initial conditions realized in the observed Universe: it
is our window to the very early universe.

\section{Adiabaticity and the CMB}

We describe the early universe by a mixture of baryons, CDM,
photons and massless neutrinos. Entropy perturbations of the
mixture are characterized by the intrinsic entropy of each
component and a contribution coming from the
mixture\cite{Kodama:84,Malik:02}. For perfect fluids, the former
vanishes, while the latter is a weighted sum over
contributions of the type
 \be
 S_{\alpha \beta} \equiv \frac{\delta_\alpha}{1 + w_\alpha} -
 \frac{\delta_\beta}{1 + w_\beta},
 \ee
for two fluids $\alpha, \beta$, where $\delta_x \equiv \delta
\rho_x / \rho_x$ is the energy contrast and $w_x \equiv
P_x/\rho_x$ is the equation of state parameter for species $x$. A
non vanishing $S_{\alpha \beta}$ corresponds to fluctuations in
the number density ratio of the two species. If these perturbations of
the entropy are such that the total density is initially unperturbed, they
 are termed isocurvature initial conditions (IC).  CDM isocurvature IC
excite a sine oscillation in the photon-baryon
fluid\cite{Durrer:01}, this corresponds (for a flat universe)
to a first peak in the temperature CMB power spectrum at a
multipole $\ell \approx 110$.

The simplest choice for IC is the one in which there is no
fluctuation in the relative number density of the species, hence
no entropy perturbations:
$$
\pert{b} = \pert{c} = \frac{3}{4}\pert{\gamma} = \frac{3}{4}
\pert{\nu} \qquad \text{(Adiabatic).}
$$
Those IC are termed adiabatic. They naturally arise from 1-field
inflationary scenarios, which have only one degree of freedom, and
therefore cannot produce entropy fluctuations. Adiabatic IC can be
described in terms of the induced curvature perturbation $\zeta$,
which in longitudinal gauge is related to the energy-density
contrast $\delta$ by
 \be
 \zeta = \left (-\frac{3}{2}\frac{\mathcal{H}^2}{k^2} + \frac{1}{3(1+w)}
 \right)
 \delta .
 \ee
Adiabatic IC excite a cosine oscillatory mode, which induces a
first peak at $\ell \approx 220$ (for a flat universe) in the CMB
angular power spectrum. The observation of the first
peak\cite{Page03} at $\ell = 220.1 \pm 0.8$ has substantially
confirmed the domination of adiabatic IC. However, a subdominant
isocurvature contribution to the prevalent adiabatic mode cannot
be excluded.

Beside AD and CDM isocurvature, the complete set of IC for a fluid
consisting of photons, neutrinos, baryons and dark matter in
general relativity consists of three more modes\cite{BMT}. These
are the baryon isocurvature mode (BI), the neutrino isocurvature
density (NID) and neutrino isocurvature velocity (NIV) modes.
Those five modes are the only regular ones, \IE they do not
diverge at early times. The NID mode can be understood as a
neutrino entropy mode, while the NIV consists of vanishing density
perturbations for all fluids but non-zero velocity perturbations
of the neutrinos. Each mode can have a different spectral index,
and cross-mode correlations can be either positive or
negative\cite{TRD1,TRD2}.

Initial conditions which represent a (anti-)correlated mixture of
the adiabatic and the CDM  isocurvature mode are obtained {\em
e.g.} in the curvaton model\cite{Enqvist,Lyth:02}.  WMAP
constraints for the curvaton model have been derived for the case
of CDM and baryons isocurvature fluctuations\cite{Gordon:02}.  The
NID mode can be generated from  perturbations of the neutrino
chemical potential\cite{Lyth2}, and bounds have recently been
derived for this case\cite{GordonMalik}. It seems more difficult
to produce a NIV mode: a working model is at present still
lacking.

\section{Model-independent constraints}

In order to test the paradigm of purely adiabatic fluctuations we
now allow for general isocurvature contributions and derive bounds
on their amplitudes and spectral index from CMB and large scale
structure data\cite{TRD2}. Although independent of any model for
the generation of perturbations, this approach has the
disadvantage of introducing many new free parameters in the
description of the power spectrum. To reduce this number somewhat,
we assume the same spectral index for all modes. Since the current
CMB data are in excellent agreement with purely adiabatic IC, it
is not surprising that there is no statistical evidence that such
extra parameters should be non-zero. Occam's razor would therefore
dictate to stick to the simplest adiabatic description, lacking
any evidence for a more complicated model. However, there is no
compelling reason why the physics of the early universe should
boil down to only one degree of freedom.

A second reason why model-independent constrains should be
regarded with care is that in any specific implementation some of
the parameters will be correlated. For instance, in the curvaton
scenario, adiabatic and residual isocurvature modes are always
totally (anti)correlated. Therefore not only the number of extra
degrees of freedom is reduced, but the parameter space of the
model is a possibly highly constrained subspace of the
model-independent parameter space.

This phenomenological approach gives useful hints on the
``stiffness'' of current data, and indeed the possibility of
accommodating isocurvature modes has been considerably reduced by
WMAP\cite{Lesgourgues:02}. However, large degeneracies between
isocurvature modes and cosmological parameters still allow for
relatively high isocurvature contribution\cite{Muonen:02,Pedro}.
The exact amount depends on the type of isocurvature mode
considered and on how many of the 5 fundamental modes are allowed
for at the same time.

\section{Initial conditions independent constraints}

With present data it is difficult to constrain at the same time
both the IC and the cosmological parameters using CMB alone. A
more powerful approach is to include data on the matter power
spectrum\cite{TRD2}, or ``priors'' on the cosmological parameters
coming from other observations\cite{Pedro}. The future high
accuracy measurements of CMB polarization will give a substantial
help in breaking degeneracies between IC. The degeneracies in the
parameter dependence of temperature and polarization are almost
orthogonal, and polarization can therefore lift ``flat
directions'' in parameter space.

\begin{figure}
\begin{center}
\epsfig{angle=270, width=0.49\linewidth, file=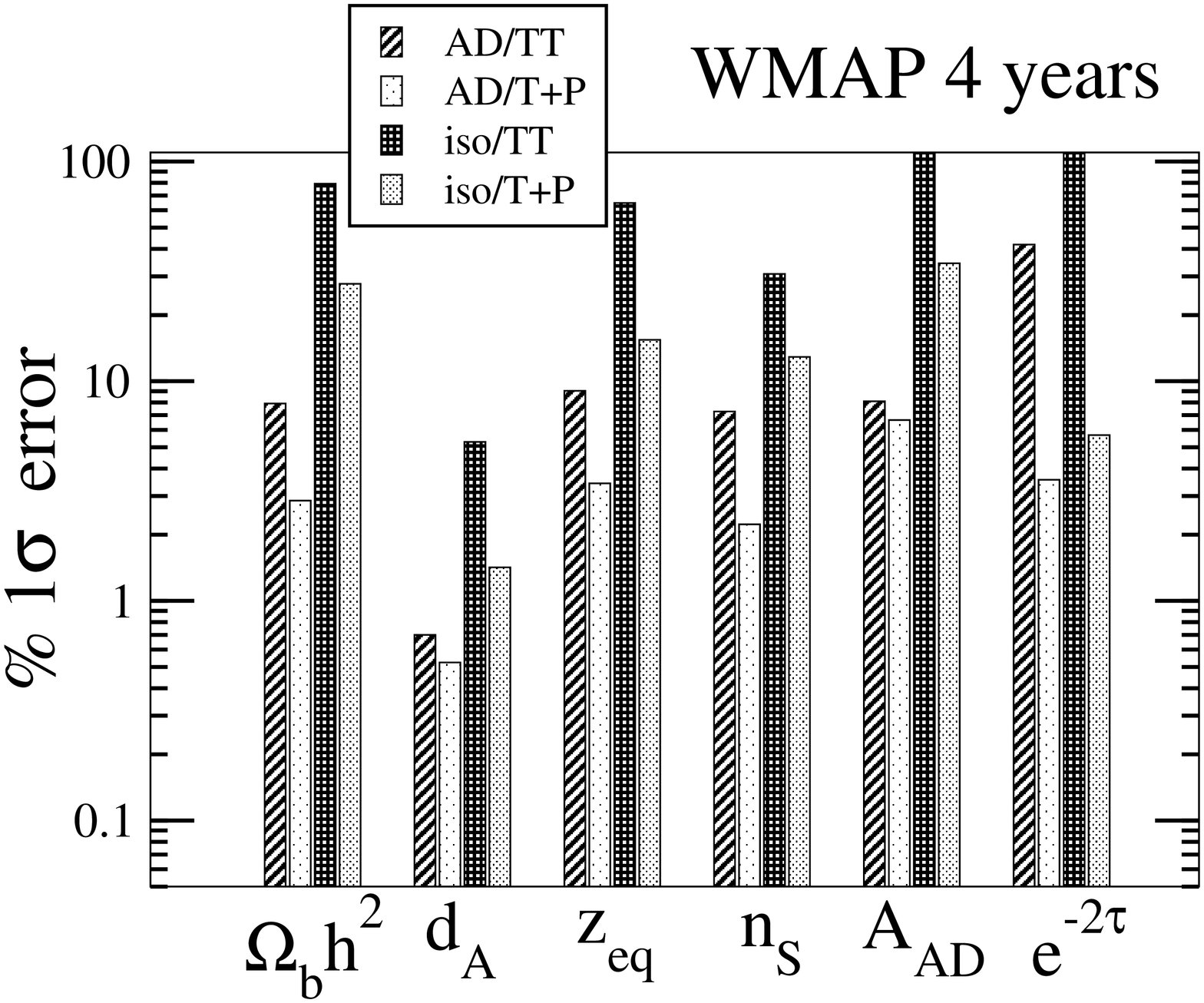}%
\hfill%
\epsfig{angle=270, width=0.49\linewidth, file=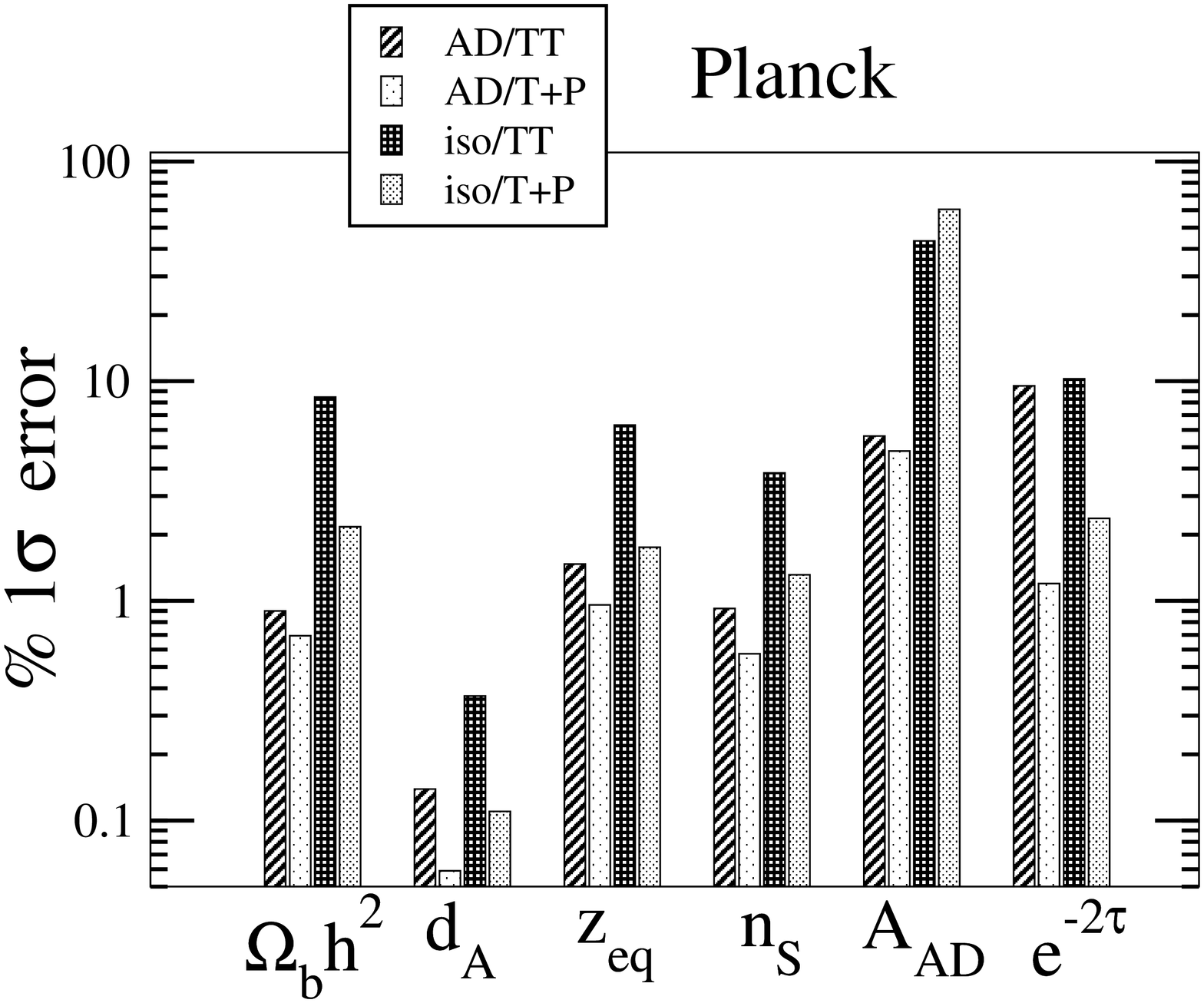}%
\end{center}
\caption{Fisher Matrix forecast for the percent $1-\sigma$ errors
on 6 quantities which are well determined by CMB alone. The left
(right) panel is a forecast for WMAP 4 years mission (Planck).
From left to right, on the abscissa axis: the baryon density,
$\Omega_b h^2$, the angular diameter distance $d_A$, the redshift
of matter-radiation equality $z_{eq}$, the scalar spectral index
$n_s$, the scalar adiabatic amplitude $A_\textrm{AD}$ and a
function of the optical depth to reionization, $\tau$. In the
legend, ``AD'' means that only adiabatic fluctuations were
included, ``iso'' means that general isocurvature modes were
included and marginalized over. ``TT'' includes temperature
information alone, ``T+P'' has temperature, E-T correlation and
E-polarization. } \label{fig:forecasts}
\end{figure}

To determine cosmological parameters independently on the IC, one
includes general isocurvature modes, and then marginalize over
them. Bucher and collaborators\cite{BMTpol} considered forecasts
for WMAP and Planck, and concluded a few years ago that such a
procedure would make it effectively impossible to constrain
parameters with meaningful precision. This result was based on a
set of cosmological parameters which has been shown to lead to
large overestimates of the expected errors\cite{Kosowsky:02}. We
have reproduced their study, using an improved Fisher Matrix
technique as in Rocha \ETAL\cite{alpha3}. In particular, we give
forecasts not for the highly degenerate directions defined by the
cosmological parameters, but rather for orthogonal combinations
which are well measured by the CMB. Along this directions
forecasts are much more reliable. The main features are shown in
Fig.~\ref{fig:forecasts}. There we plot the expected $1-\sigma$
error in percent for 6 quantities which are directly probed by the
CMB (see figure caption).

For WMAP, marginalization over general initial conditions will
indeed give errors which for all quantities will be roughly a
factor 10 larger than in the purely adiabatic case, when
temperature information alone is considered (cf first and third
bar in the left panel). When the full polarization information is
included, the errors will still be within approximately 10 to 30\%
even in the general isocurvature scenario. From the right panel,
we deduce that for the Planck experiment\cite{Planck} the worsening of
the errors will be much less if the high quality polarization information
is included. Roughly, including isocurvature modes we expect
errors which are larger than in the adiabatic case by about a
factor of 2, but mostly still within the few percent accuracy.

This shows that the CMB alone will be able to provide high
precision cosmology even if the strong assumption of purely
adiabatic initial conditions will be relaxed. Combining CMB
results with other observation which independently constrain the
cosmological parameters, will enable us to fully open this window
to the mysterious epoch of the very early universe.

\end{document}